\documentclass[aps,preprint]{revtex4}
\usepackage{graphicx}
\usepackage{amsmath}

\begin{document}
%\draft

% You should use BibTeX and revtex.bst for references
%\bibliographystyle{apsrev}

%\preprint{}

\title{Laminar-Turbulent Transition: The change of the flow state temperature
with the Reynolds number}
% Optional argument for running titles on pages
%\title[]{}


% \affiliation command. The \affiliation command MUST follow the

\author{Sergei F. Chekmarev}

\affiliation {Institute of Thermophysics, 630090 Novosibirsk,
Russia, and \\Department of Physics, Novosibirsk State University,
630090 Novosibirsk, Russia}

%\affiliation{}


\date{\today}

\begin{abstract}
Using the previously developed model to describe laminar/turbulent states of a viscous
fluid flow, which treats the flow as a collection of coherent structures of various size
(Chekmarev, Chaos, 2013, 013144), the statistical temperature of the flow state is
determined as a function of the Reynolds number. It is shown that at small Reynolds
numbers, associated with laminar states, the temperature is positive, while at large
Reynolds numbers, associated with turbulent states, it is negative. At intermediate
Reynolds numbers, the temperature changes from positive to negative as the size of the
coherent structures increases, similar to what was predicted by Onsager for a system of
parallel point-vortices in an inviscid fluid. It is also shown that in the range of
intermediate Reynolds numbers the temperature exhibits a power-law divergence
characteristic of second-order phase transitions.
\end{abstract}
\pacs{}

%\maketitle must follow title, authors, abstract and PACS
\maketitle


\section{Introduction}
In his famous work on statistical hydrodynamics \cite{Onsager49}, Onsager has predicted
that a system of vortices can have negative temperature states, which can be interpreted
as a clustering of vortices of the same sign. More specifically, Onsager considered a
microcanonical ensemble of parallel point-vortices in an incompressible inviscid fluid
confined to a finite region of physical space and calculated the entropy of the system
$S$ as a function of its energy $E$. Taking into account that the coordinates of point
vortices obey Hamiltonian equations and thus are canonically conjugate, he concluded that
the phase volume of the system $\Phi(E)$ is also finite, so that the density of states
$\Gamma(E)=\Phi '(E)$ should have a maximum value at some (critical) energy $E_{\mathrm
m}$. Correspondingly, the statistical temperature $T=(dS/dE)^{-1}$, where $S(E)=\ln
\Gamma(E)$ is the entropy, should be positive at $E < E_{\mathrm m}$ and negative at $E
> E_{\mathrm m}$. Using these arguments, Onsager has concluded that "the large compound
vortices formed in this manner will remain as the only conspicuous features of the
motion; because the weaker vortices, free to roam practically at random, will yield
rather erratic and disorganised contributions to the flow". This discovery of the
negative temperature states has given rise to a number of theoretical considerations,
where Onsager's arguments were revisited, discussed and elaborated
\cite{JoyceMontgomery73,EdwardsTaylor74,MontgomeryJoyce74,PointinLundgren76,LundgrenPointin77,
KraichnanMontgomery80,FrohlichRuelle82,Miller90,Chorin91,EyinkSpohn93,JungMorrisonSwinney06,
Chavanis12} (for review, see Eyink and Sreenivasan \cite{EyinkSreenivasan06} and Campa et
al. \cite{CampaDauxoisRuffo09}). In particular, it was shown that the key assumption of a
finite region of physical space is not crucial because the accessible phase space is
restricted by a constant of motion, which is an analog of enstrophy
\cite{LundgrenPointin77,EyinkSpohn93}.

Recently, we have proposed a statistical model for a fluid flow, which considers the flow
as a collection of spatially localized (coherent) structures and uses the principle of
maximum entropy to determine the most probable size distribution of the structures
\cite{Chekmarev13}. A principal difference of this model from the previous models in
which a similar approach was used to describe inviscid flows (for a review of such
models, see, e.g., Refs. \onlinecite{EyinkSreenivasan06} or \onlinecite{Chekmarev13}) is
that the structures are assumed to be composed of elementary cells in which the behavior
of the particles (atoms or molecules) is uncorrelated. The characteristic size of
elementary phase space volumes corresponding to these cells is determined by the
kinematic viscosity of the fluid, which makes it possible to introduce the Reynolds
number and thus to extend the statistical description of the flow to viscous fluids. The
model successfully describes the transition to turbulence at large Reynolds numbers and
some other characteristic properties of turbulent flows \cite{Chekmarev13}. Since the
dependence of the energy and entropy of the flow on the Reynolds number is available in
this model, the statistical temperature of the flow state can be calculated as a function
the Reynolds number. In the present paper we show that at low Reynolds numbers,
associated with laminar motion, the temperature is positive, and at high Reynolds
numbers, associated with turbulent motion, it is negative. At intermediate Reynolds
numbers, representing the transition range, the temperature changes from positive to
negative as the structure size exceeds some critical value, similar to what was
discovered by Onsager for the system of parallel point-vortices in a inviscid fluid. We
also show that in the range of intermediate Reynolds numbers the temperature exhibits a
power-law divergence characteristic of second-order phase transitions.

The paper is organized as follows. Section II briefly describes the statistical model of
the flow that was developed in Ref. \onlinecite{Chekmarev13} and serves as an
introduction to Sect. III. Section III introduces the statistical temperature of the flow
states and presents the results of the study and their analysis, including the analogy
with second-order phase transitions. Section IV contains concluding remarks.

\section{The model}
The model is briefly described as follows (for details, see Ref. \cite{Chekmarev13}). Let
a fluid flow be represented by a system of $N$ identical particles (atoms or molecules)
which are placed in a volume of physical space $V$. Assume that the particles can form
spatially localized structures of $N_i$ particles ($1 \le N_i \le N$), with the number of
such ($i$) structures being $M_i$. These structures are viewed as coherent structures,
i.e., the structures in which the constituting particles execute a concerted motion on
the structure scale. Since such concerted motion should break down at the microscale
level, each $i$ structure can be divided into $N_{i}/n$ elementary "cells" of volume $v$
(each of $n$ particles), in which the behavior of the particles is uncorrelated. Because
the particles are identical, to each elementary cell there corresponds a 6-dimensional
elementary volume of single-particle phase space, in which the state of the particles is
uncertain. For a gas fluid, the velocity and linear scales of such a volume are,
respectively, molecular thermal velocity $c$ and mean free path $\lambda$ (Kusukawa
\cite{Kusukawa51}). According to the kinetic theory of gases $ c \lambda \sim \nu$, where
$\nu$ is the kinematic viscosity (see, e.g., Ferziger and Kaper \cite{FerzigerKaper}),
i.e., the characteristic linear size of the elementary phase volume is determined by the
kinematic viscosity \cite{Kusukawa51}. For a liquid fluid, the linear size of the
elementary volume can be taken to be $c^2 \tau$, where $c^2$ is the fluctuation of the
(molecular) kinetic energy per unit mass, and $\tau$ is the mean residence time of the
molecule in a settled state, so that a similar estimate is valid, $c^2 \tau \sim \nu$
(e.g., Frenkel \cite{Frenkel}). Alternatively, the space and velocity scales of the
volume can be associated with the corresponding Kolmogorov microscales
\cite{Chekmarev13}, i.e., $\eta =(\nu^{3}/\epsilon)^{1/4}$ and $v_{\eta}=(\nu
\epsilon)^{1/4}$ , respectively ($\epsilon$ is the rate of dissipation per unit mass)
\cite{Kolmogorov41a,LandauLifshitz_fluid_mech}.

The given system of coherent structures is characterized by two parameters, which are the
total number of particles (atoms or molecules) in the system $N$ and the total energy
$E$, i.e., the corresponding ensemble of the systems is a {\it microcanonical} ensemble.
We assume that the total energy can be written as $E=\sum_{i}M_{i}E_{i}$, where $E_{i}$
is the kinetic energy of the concerted motion of particles in $i$ structure. This
definition is based on two assumptions: i) the concerted motion of particles in each
local region of the physical space is associated with one of the coherent structures and
contributes to its kinetic energy, and ii) the potential energy of interaction of
coherent structures is either negligible in comparison to the kinetic energy of the
system or, if not, can be considered as a part of its kinetic energy, as, e.g., for the
system of two-dimensional point vortices in inviscid fluid (Batchelor \cite{Batchelor}),
which was studied by Onsager \cite{Onsager49}. Given the above definition of the total
energy, the collection of the coherent structures can be treated as an ideal gas system,
i.e., a system in which the potential energy is negligible and the total energy is
additive (Landau and Lifshitz \cite{LandauLifshitz_stat_phys}). It is important that the
ideal gas approximation does not imply that the interaction in the system is absent at
all. One example, which was used by Ruelle \cite{Ruelle12} to draw an analogy to the
turbulence cascade, is the heat conductivity in a rarefied gas: while the gas is
considered to be an ideal gas, the heat transfer exists due to the energy exchange
between the molecules in the course of their short-range collisions. This example is also
useful in another respect. Although the process of the heat transfer is a non-equilibrium
process, the molecular velocity distribution function, which corresponds to the the
Fourier law, is just slightly different from the equilibrium (Maxwell) distribution, as a
first-order perturbation of the latter (e.g., Ferziger and Kaper \cite{FerzigerKaper}).
Therefore, if the statistical properties of the system is of interest, the equilibrium
distribution function can be considered as a zero-order approximation.

Every distribution of particles (atoms or molecules) among the coherent structures that
satisfies the conditions $\sum_{i}N_{i}M_{i}=N$ and $E=\sum_{i}M_{i}E_{i}$ presents a
{\it microstate} of the system at given $N$ and $E$, and the number of such microstates
$\Gamma(N,E)$ plays a role of the {\it density} of states. Taking into account that the
particles are identical, so that the permutations of the particles in the elementary
cells, the permutations of the elementary cells in the coherent structures, and the
permutations of the coherent structures themselves do not lead to new states, we have
\begin{equation}\label{eq0}
\Gamma(N,E)=\frac{N!}{\prod_{i}\left[(n!)^{N_{i}/n}(N_{i}/n)!\right]^{M_i}M_{i}!}
\end{equation}
or, with the Stirling approximation $x!\approx (x/e)^x$ to be applicable
\begin{equation}\label{eq1}
\Gamma(N,E)=\frac{(N/n)^{N}e^{N/n}}{\prod_{i}(N_{i}/n)^{N_{i}M_{i}/n}(M_{i}/e)^{M_{i}}}
\end{equation}
Using Eq. (\ref{eq1}), it is possible to calculate the most probable size distribution of
the structures, $\tilde{M}_{i}=\tilde{M}_{i}(N_{i})$, which maximizes the entropy
$S(N,E)=\ln \Gamma(N,S)$ (the Boltzmann constant is set to unity) and determines the
observational {\it macrostate} of the system. The variation of the entropy functional
with respect to $M_{i}$ at two conservation conditions $\sum_{i}M_{i}N_{i}=N$ and
$\sum_{i}M_{i}E_{i}=E$ yields
\begin{equation}\label{eq2}
\tilde{M}_{i}%
=\left(\frac{N_{i}}{n}\right)^{-\frac{N_{i}}{n}}e^{\alpha' N_{i}+\beta'E_{i}}
\end{equation}
where and $\alpha'$ and $\beta'$ are the constants depending on $N$ and $E$ (the Lagrange
multipliers).

\begin{figure}\centering%
\resizebox{0.6\linewidth}{!}{ \includegraphics*{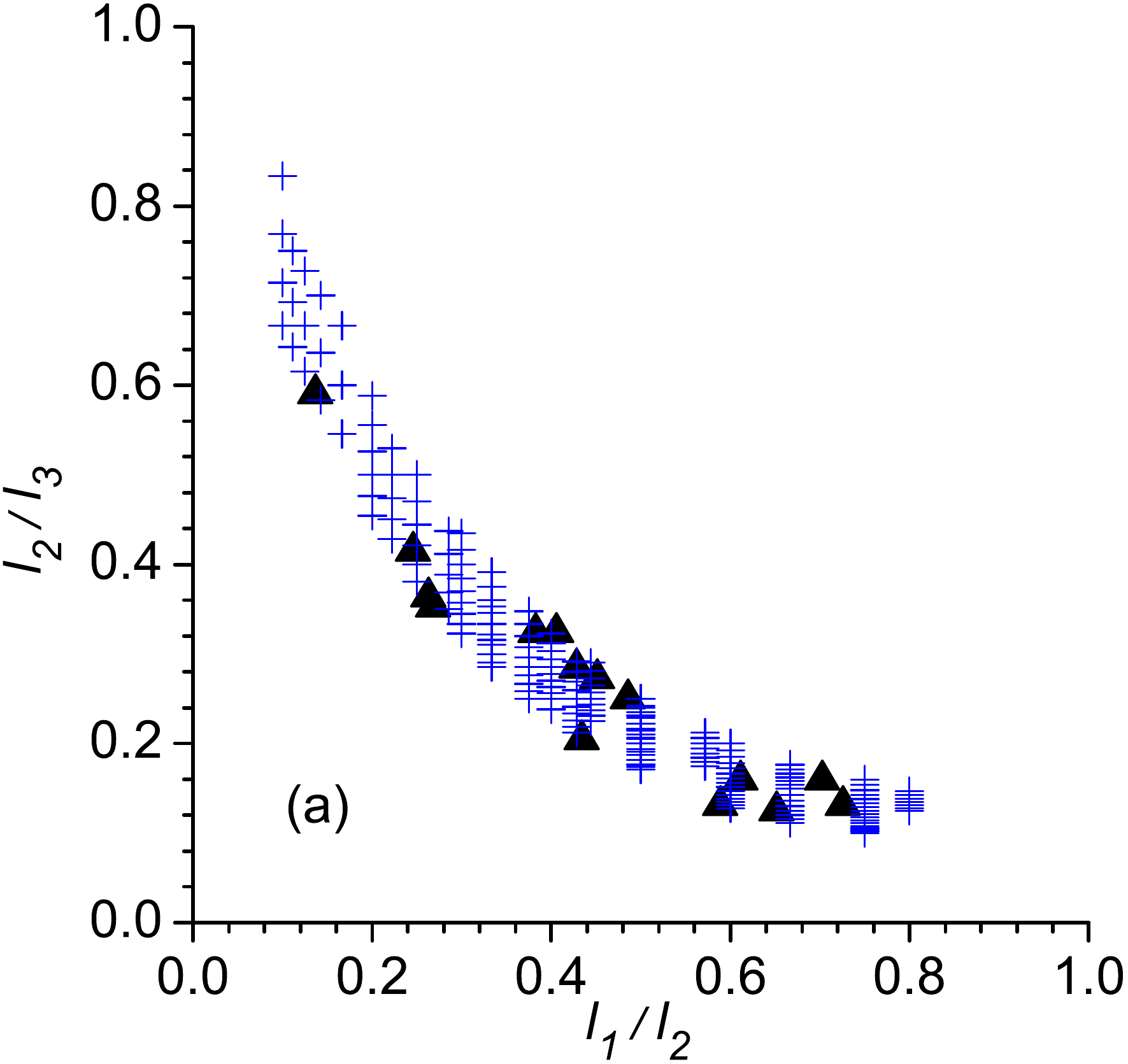}}%
\vfill
\resizebox{0.6\linewidth}{!}{ \includegraphics*{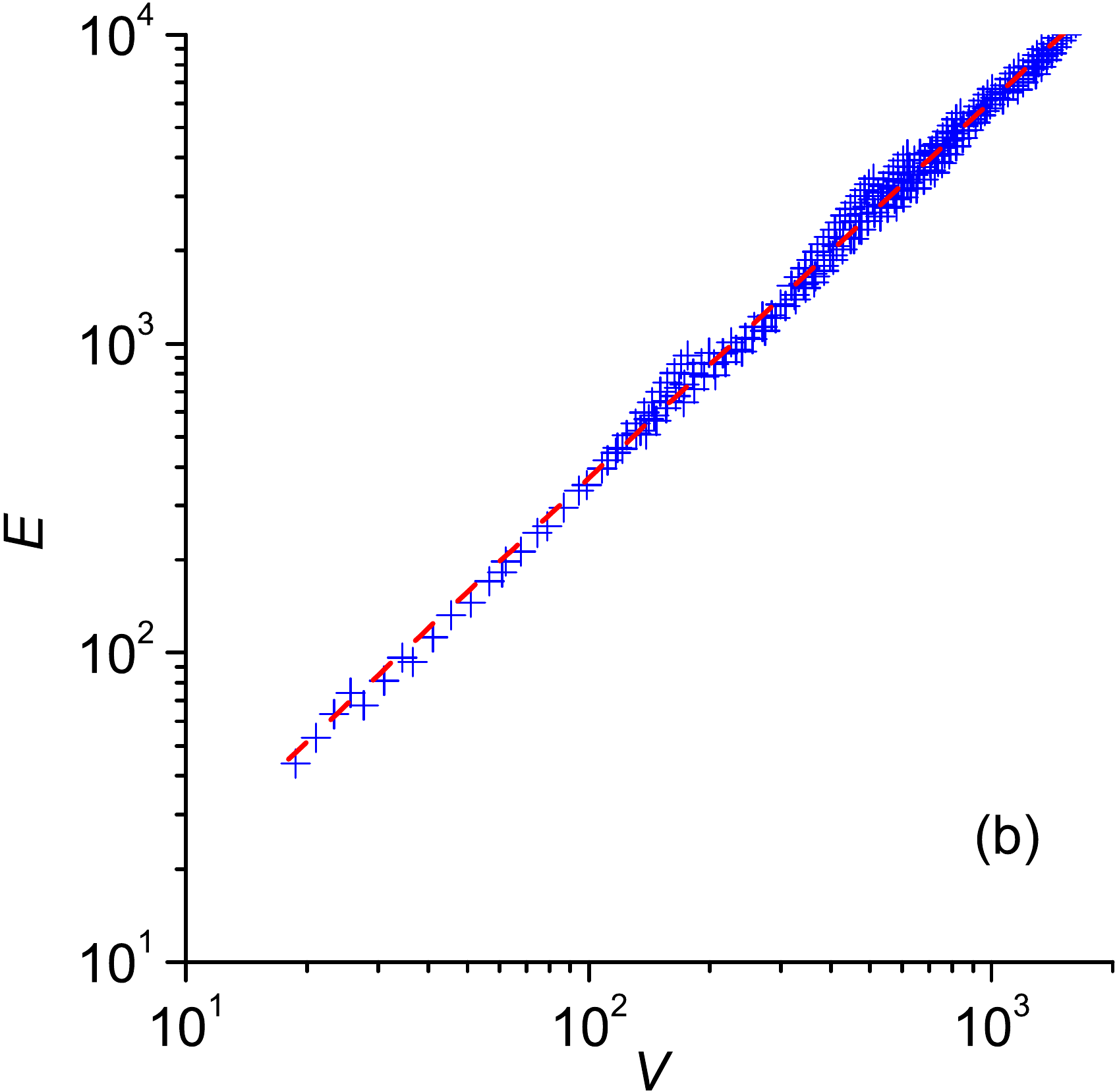}}%
\caption{(Color online) Properties of the coherent structures. (a) The aspect ratios of
the structures: the triangled correspond to Ref. \onlinecite{MoisyJimenez04}, and the
crosses to the present work. (b) The kinetic energy of the structure versus its volume:
the crosses correspond to points shown by crosses in panel (a), and the dashed line shows
the fitting of the data to the power function $E \sim V^{1.23}$.} \label{aspect_rations}
\end{figure}

To specify the dependence of $E_{i}$ on $N_{i}$, we followed Kolmogorov
\cite{Kolmogorov41a}, i.e., we assumed that the energy of concerted motion of particles
increased with distance as $e(l) \sim l^{2h}$, where $h=1/3$. Since the coherent
structures may be different in form, the law of dependence of $E_{i}$ on $N_{i}$ can also
be different. To estimate the law, we calculated the kinetic energy in a parallelepiped
with sides $L_{1}$, $L_{2}$ and $L_{3}$ as $E(L_{1},L_{2},L_{3}) \sim
\int_{-L_{1}/2}^{L_{1}/2}\int_{-L_{2}/2}^{L_{2}/2}\int_{-L_{3}/2}^{L_{3}/2} r^{2h}
dl_{1}dl_{2}dl_{3}$, where $r^2=l_{1}^2+l_{2}^2+l_{3}^2$, and related the obtained energy
to the parallelepiped volume $V=L_{1}L_{2}L_{3}$. The results of the calculations are
presented in Fig. \ref{aspect_rations}. Panel {\bf{a}} shows typical coherent structures
that were observed by Moisy and Jim\'{e}nez in direct numerical simulations of isotropic
turbulence \cite{MoisyJimenez04}, who characterized the structures by the aspect ratios
$L_{1}/L_{2}$ and $L_{2}/L_{3}$ (solid triangles). There are also shown the corresponding
structures we chosen to calculate the dependence of the kinetic energy on the structure
volume (crosses). Panel {\bf{b}} of Fig. \ref{aspect_rations} depicts the obtained
dependence of $E_{i}$ on $V_{i}$ for these structures. As is seen, it is well fitted by
the function $E \sim V^{\gamma}$ with $\gamma \approx 1.23$, which is close to
$\gamma=2h/3+1=11/9 \approx 1.22$ characteristic of the spherical structures
\cite{Chekmarev13}. The same exponent $\gamma=11/9$ was used by Ruelle \cite{Ruelle12},
who, instead of introducing the coherent structures, divided the physical space in cubes
in which the kinetic energy of fluctuations was assumed to obey the Kolmogorov theory
\cite{Kolmogorov41a}. In the present paper, the calculations are made with $\gamma=11/9$.
Assuming the density of the fluid to be constant, we thus have $E_{i} \sim
N_{i}^{\gamma}$.

Strictly speaking, the Kolmogorov relation assumes the inertial interval of scales $\eta
\ll l \ll L$, where $\eta$ and $L$ are the dissipation and external scales, respectively.
However, for rough estimates, the range of its validity can be extended to the lower and
upper bounds, i.e., to $\eta \le l \le L$ \cite{Chekmarev13}. Moreover, with the
arguments given by Landau and Lifshitz \cite{LandauLifshitz_fluid_mech}, the lower bound
can be associated with laminar motion \cite{Chekmarev13}. According to this, we associate
the laminar state with the structures that are comparable in size with the elementary
cell, i.e., $N_i/n \sim 1$, and the turbulent state with those that contain many
elementary cells, i.e., $N_i/n \gg 1$.

After substituting the dependence $E_i$ on $N_i$ into Eq. (\ref{eq2}), we eventually have
\begin{equation}\label{eq3}
\tilde{M}_{i}=q_{i}^{-q_{i}}e^{\alpha q_{i}+\beta q_{i}^{\gamma}}
\end{equation}
where $q_{i}=N_{i}/n$ is the number of the elementary cells in $i$ structure, and
$\alpha$ and $\beta$ are new constants that should be determined from the equations of
conservation of the total number of particles and kinetic energy in the form
\begin{equation}\label{eq3_1}
N/n=\sum_{i}\tilde{M}_{i}q_{i}
\end{equation}
and
\begin{equation}\label{eq3_2}
E/n^{\gamma}=\sum_{i}\tilde{M}_{i}q_{i}^{\gamma}
\end{equation}
It is also possible, and is more convenient, to vary $\alpha$ and $\beta$ to obtain $N/n$
and $E/n^{\gamma}$ as functions of $\alpha$ and $\beta$ \cite{Chekmarev13}.

The parameter $N/n$ can be associated with the Reynolds number ${\mathrm
{Re}}_{L}=WL/\nu$, where $W$ and $L$ are, respectively, the velocity and linear scales
characterizing the system as a whole. As has been mentioned, the characteristic size of
the elementary phase volume is determined by the kinematic viscosity, so that the number
of particles in the volume $n \sim \nu^3$. Correspondingly, the total number of particles
$N$ can be considered to be proportional to the total phase volume for the system, i.e.,
$N \sim (WL)^3$. Then $N/n \sim (WL/\nu)^3 ={\mathrm {Re}}_{L}^3$, or assuming for
simplicity that the coefficient of proportionality is equal to 1
\begin{equation}\label{eq4}
{\mathrm {Re}}_{L}=(N/n)^{1/3}
\end{equation}

\begin{figure}\centering%
\resizebox{0.8\linewidth}{!}{ \includegraphics*{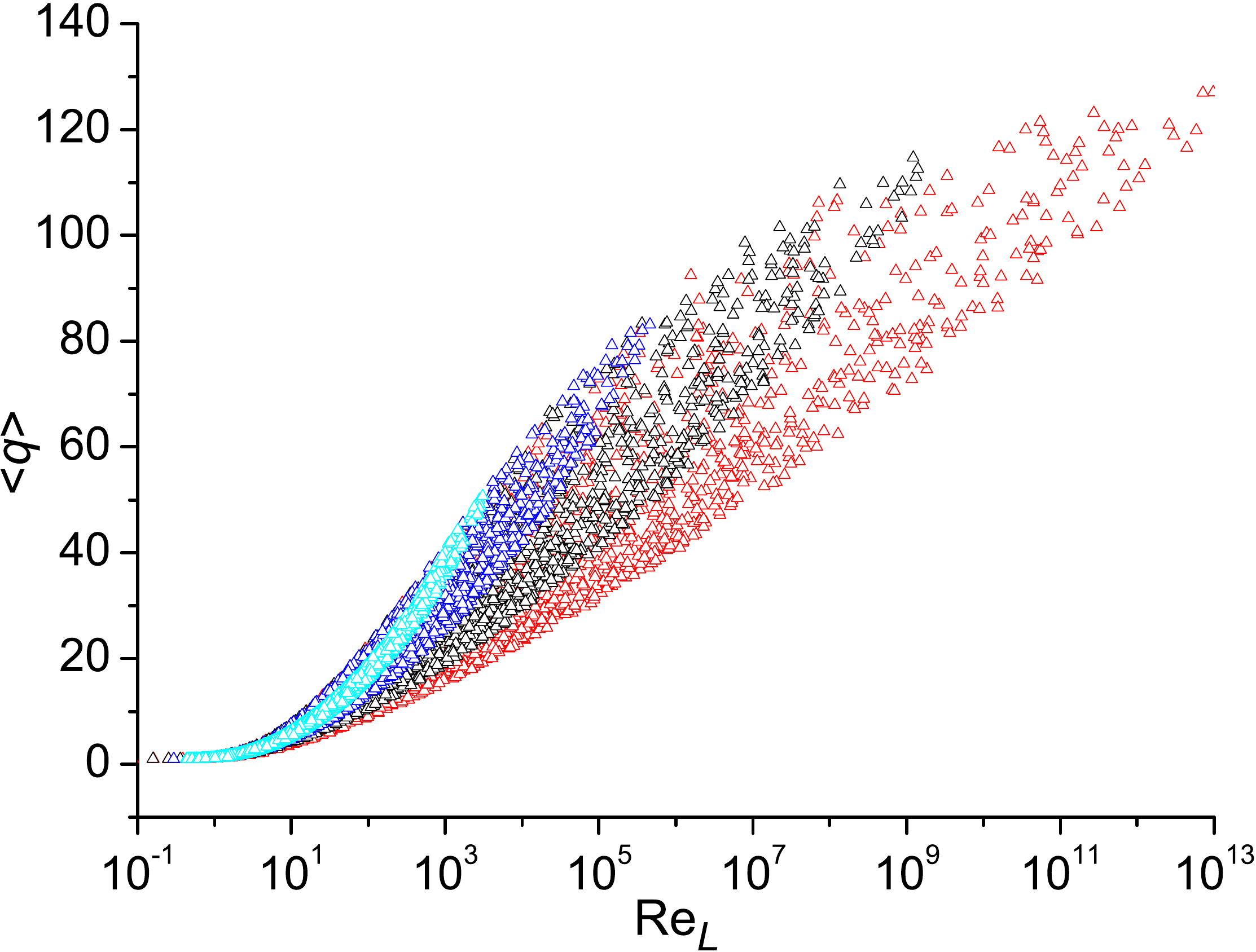}}%
\caption{(Color online) Average number of the elementary cells in the coherent structures
as a function of the Reynolds number. The constants $\alpha$ and $\beta$ vary from -1.3
to 1.3 (cyan labels), -2.0 to 2.0 (blue), -3.0 to 3.0 (black), and -4.0 to 4.0 (red).}
\label{q(re)}
\end{figure}

The distribution given by Eq. (\ref{eq3}) is drastically different from the conventional
Gibbs-Boltzmann distribution for a many-particle system (see, e.g., Landau and Lifshitz
\cite{LandauLifshitz_stat_phys}). In particular, none of the Lagrange multipliers
$\alpha$ and $\beta$ can be associated with temperature. As has been shown in our
previous paper \cite{Chekmarev13}, these multipliers determine the range of the Reynolds
numbers where a turbulent state can be expected. To illustrate this, Fig. \ref{q(re)}
shows the average number of elementary cells in the coherent structures $\langle
q\rangle=\sum_{i}q_{i}\tilde{M}_{i}/\sum_{i}\tilde{M}_{i}$ as a function of the Reynolds
number ${\mathrm {Re}}_{L}$ determined via Eqs. (\ref{eq3_1}) and (\ref{eq4}). The number
of the elementary cells $q_i$ varied from 1 to $q_{\mathrm{max}}=200$; a variation of
$q_{\mathrm{max}}$ does not change the overall picture, except that the ranges of
variation of $\langle q\rangle$ and ${\mathrm {Re}}_{L}$ extend to larger values of these
quantities as $q_{\mathrm{max}}$ increases \cite{Chekmarev13}. The values of $\alpha$ and
$\beta$ were randomly chosen from the uniform distributions within $\alpha_{\mathrm{min}}
\leq \alpha \leq \alpha_{\mathrm{max}}$ and $\beta_{\mathrm{min}} \leq \beta \leq
\beta_{\mathrm{max}}$, respectively (for $10^5$ samples in each case). As was indicated
in Ref. \onlinecite{Chekmarev13}, not every combination of $\alpha$ and $\beta$ gives a
physically reasonable bell-shaped distribution \cite{JimenezWraySaffmanRogallo93}, i.e.,
for some combinations, the fraction of which is typically within several percentages,
$\tilde{M}_{i}$ does not vanish at $q \rightarrow q_{\mathrm{max}}$. In Fig. \ref{q(re)},
the points with such "wrong" $\alpha / \beta$ combinations are excluded.

According to the present model, the points of Fig. \ref{q(re)} for which $\langle
q\rangle \sim 1$ should be associated with a laminar state, and those for which $\langle
q\rangle \gg 1$ with a turbulent state. A boundary that seemingly separates these states
lies at a Reynolds number ${\mathrm {Re}}^{\star}_{L}$, which is between ${\mathrm
{Re}_{L}} \sim 10$ and ${\mathrm {Re}_{L}} \sim 10^2$. The positions of the lower and
upper boundaries of the manifold of points representing $\langle q\rangle$ as a function
of ${\mathrm {Re}}_{L}$ are determined by the maximum values of $\beta$ and $\alpha$,
respectively \cite{Chekmarev13}. The lower boundary shifts to lower values of ${\mathrm
{Re}}_{L}$ as $\beta$ increases, and the upper boundary shifts to larger values of
${\mathrm {Re}}_{L}$ as $\alpha$ increases, specifically as ${\mathrm {Re}}_{L} \sim
\exp(-0.15 \beta\langle q\rangle)$ and ${\mathrm {Re}}_{L} \sim \exp(0.06 \alpha\langle
q\rangle)$, respectively. However, while the upper boundary shifts unlimitedly, the lower
boundary changes its position until the value $\beta \approx 1.5$ is reached, after that
it "freezes" \cite{Chekmarev13} (see also Fig. \ref{q(re)}). It is significant that the
dependence of $\langle q \rangle$ on ${\mathrm {Re}}_{L}$ is not unique, i.e., the state
of the flow is not solely determined by the Reynolds number; rather, as is well-known, it
can depend on the type of the flow, the inlet conditions, the flow environment, etc.
\cite{Lesieur}. In this respect, it is important that for large values of $\alpha$, the
states with $\langle q \rangle \sim 1$ are present at the Reynolds numbers far above its
"critical" value ${\mathrm {Re}}^{\star}_{L}$; for example, at $\alpha=50$ the values of
$\langle q \rangle \approx 3$ are found up to ${\mathrm {Re}}_{L} \sim 1 \times 10^4$.
The model thus predicts that the flow should not only be laminar at ${\mathrm {Re}}_{L} <
{\mathrm {Re}}^{\star}_{L}$ but can also remain laminar at ${\mathrm {Re}}_{L} \gg
{\mathrm {Re}}^{\star}_{L}$, as it was observed, e.g., for pipe flows by Reynolds
\cite{Reynolds1883} and confirmed in many subsequent studies
\cite{DarbyshireMullin95,Pfenninger61,EckhardtSchneiderHofWesterweel07}. We note that the
specific properties of the coherent structures, i.e., the structure shape and the
velocity distribution inside the structure, do not seem to be of critical importance
here. In particular, the structures in the form of the Burgers vortices of finite length
(to mimic "worms") lead to the dependence of the structure size on the Reynolds number
that is qualitatively similar to that in Fig. \ref{q(re)} \cite{Chekmarev13}.

A highly nontrivial property of the present system, which distinguishes it from the
conventional many-particle systems in statistical physics
\cite{LandauLifshitz_stat_phys}, is that the entropy of the system is non-additive. For
example, if we divide the given system of $N$ particles into $k$ identical and
noninteracting subsystems, the density states of the composite system will be
\begin{equation}\label{eq4_1}
\Gamma_{\Sigma}=\frac{[(N/k)!]^{k}}%
{\{\prod_{i}\left[(n!)^{N_{i}/n}(N_{i}/n)!\right]^{L_i}L_{i}!\}^{k}}
\end{equation}
where $L_{i}=M_{i}/k$. Comparison of this equation with Eq. (\ref{eq0}) yields
\begin{equation}\label{eq4_2}
S_{\Sigma}-S=\ln(\Gamma_{\Sigma}/\Gamma)=\sum_{i}(1-N_{i})M_{i}\ln k \approx -N \ln k%
=-n{\mathrm{Re}}_{L}^3 \ln k < 0
\end{equation}
where $S_{\Sigma}$ and $S$ are the entropies of the composite and original systems,
respectively. The non-additivity of entropy is consistent with the fact that the system
of particles representing a fluid flow cannot be divided into parts without a loss of its
essential properties. According to Eq. (\ref{eq4}), the division of the system into
subsystems will reduce the Reynolds number characterizing a subsystem (in the above
example, in $k^{1/3}$ times), so that the resulting Reynolds number for the subsystem
will not correspond to the flow state in the subsystem, which is assumed to be the same
as in the total system. In other words, the condition that the Reynolds number is
determined by the number of particles in the system [Eq. (\ref{eq4})], originates a
virtual interaction of the coherent structures that makes unfavorable the separation of
the flow into domains at equilibrium. As can be seen from Eq. (\ref{eq0}), the entropy of
the system is additive only at the level of the coherent structures, i.e., each
collection of the structures of a specific size gives an additive contribution to the
system entropy.

According to Eq. (\ref{eq0}), the density of states $\Gamma$ and, correspondingly, the
entropy $S=\ln \Gamma$, assume that the particles (atoms or molecules) are identical, and
an elementary volume of the phase space exists in which the state of the particles is
uncertain. The same assumptions play a key role in the derivation of the Navier-Stokes
equation from the the Liouville equation, and thus they are implicitly present in the
Navier-Stokes equation \cite{Chekmarev13} (for the procedure of the derivation of the
Navier-Stokes equation for a gas fluid from the Liouville equation see, e.g., Ferziger
and Kaper \cite{FerzigerKaper}). More specifically, assuming the particles to be
identical (the first assumption of the present model), the Liouville equation for the
many-particle distribution function is reduced to a kinetic equation which contains
single-particle and two-particle distribution functions (the
Bogoliubov-Born-Green-Kirkwood-Yvon hierarchy). Then, using the molecular chaos
assumption (which corresponds to the second assumption of the model), this equation is
closed by expressing the two-particle distribution function through the single-particle
one to yield the Boltzmann equation. The Navier-Stokes equation is obtained from the
Boltzmann equation in the limit of small Knudsen number $\mathrm{Kn}=\lambda/L$, as the
first-order correction in $\mathrm{Kn}$ to the Euler equation, which describes the gas
motion in a state of local equilibrium (the Chapman-Enskog expansion). Therefore,
although the Navier-Stokes equation is incomparably rich in the description of the
laminar/turbulent transition, giving a dynamic picture of the process, it is reasonable
to expect that the statistical properties of the flow calculated on the basis of the
Navier-Stokes equation are, in a considerable degree, due to the above assumptions
involved in the derivation of the Navier-Stokes equation (they can be considered as a
hidden statistical content of the Navier-Stokes equation).

\section{Temperatures of the flow states: Results and discussion}
\begin{figure}\centering%
\resizebox{0.8\linewidth}{!}{ \includegraphics*{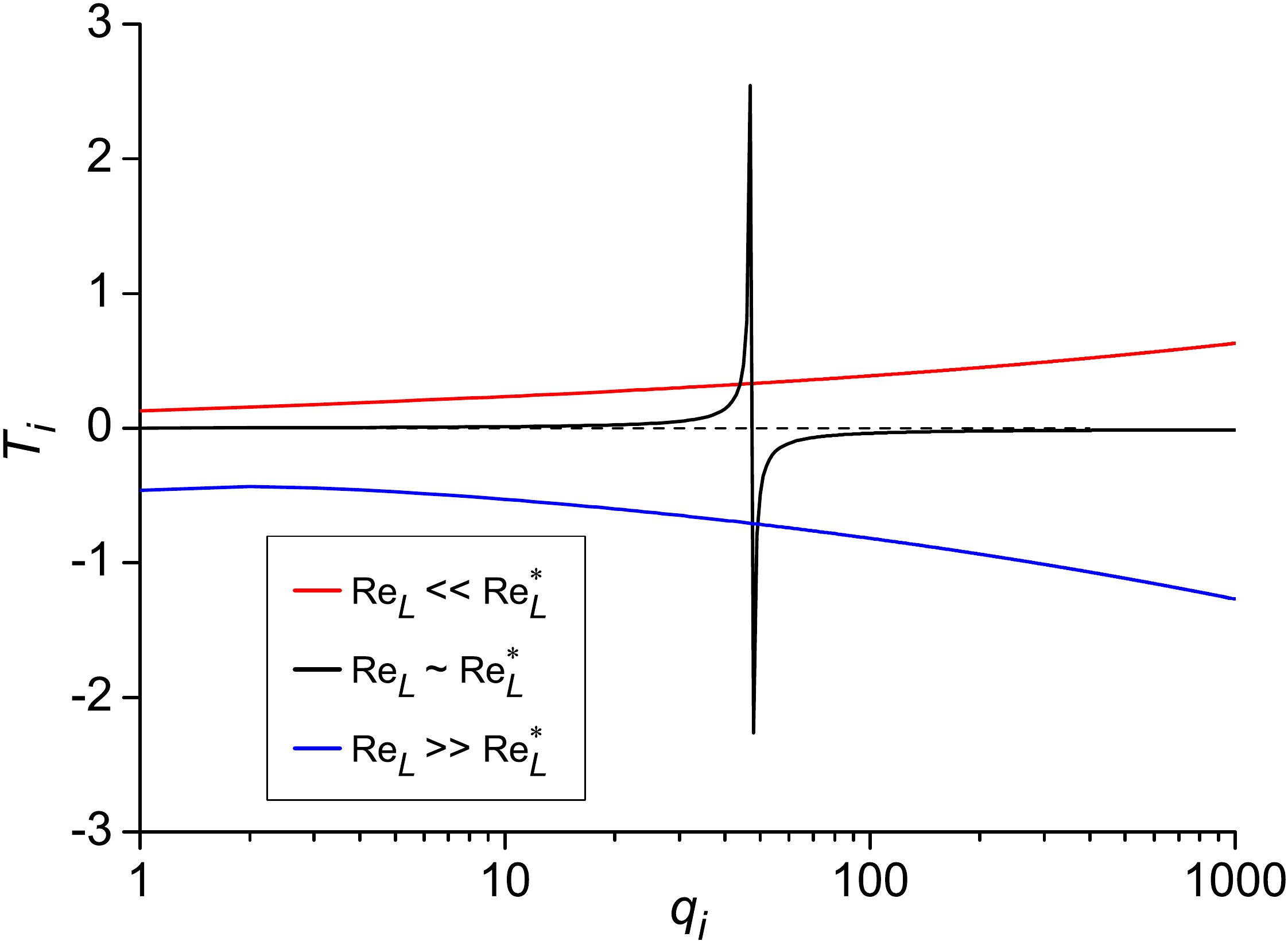}}%
\caption{(Color online) Characteristic dependencies of the flow state temperature on the
structure size. The red, black and blue curves are for the small, intermediate and large
Reynolds numbers, respectively. Specifically, the curves correspond to ${\mathrm
{Re}}_{L}=0.2$ (red), ${\mathrm {Re}}_{L}=1$ (black), and ${\mathrm {Re}}_{L}=100$
(blue). For illustration purpose, the temperature given by the black curve is decreased
by 50 times. } \label{t(q)}
\end{figure}

\begin{figure}\centering%
\resizebox{0.8\linewidth}{!}{ \includegraphics*{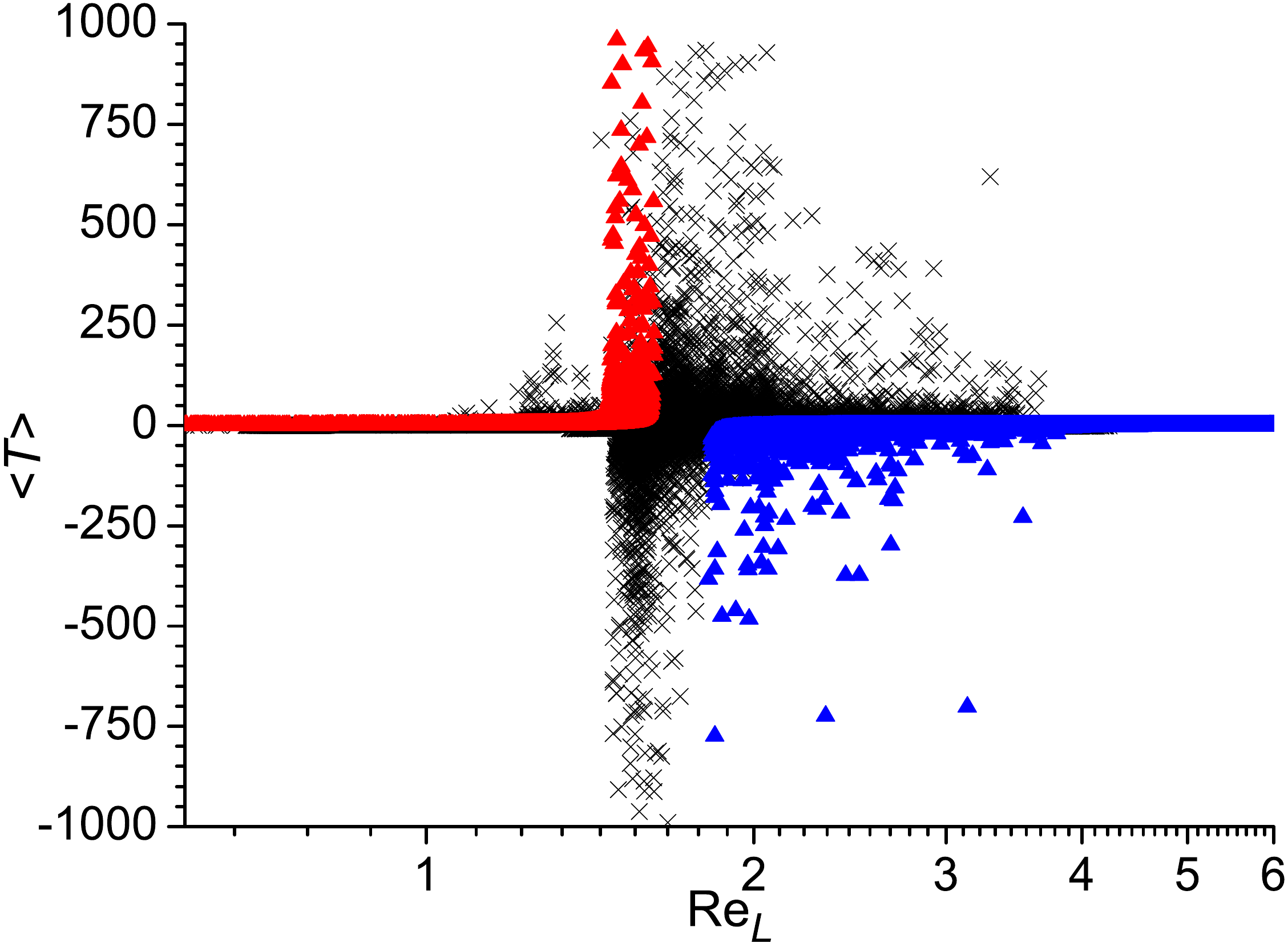}}%
\caption{(Color online) Average temperatures of the flow states as a function of the
Reynolds number. The red triangles correspond to the laminar state, the blue triangles to
the turbulent state, and the crosses to the transition states. To generate the values of
$T_i$ to be averaged, the constants $\alpha$ and $\beta$ varied from -7.0 to 7.0. For
purpose of illustration, not all points are shown: the temperatures are restricted by the
value of $|<T>|=10^3$ (though they can be as large as $|<T>| \sim 10^4$), and only the
points whose weight in each subset of the points (represented by the red and blue
triangles, and the crosses) are not less than $1 \times 10^{-6}$.} \label{t(re)}
\end{figure}

\begin{figure}\centering%
\resizebox{0.8\linewidth}{!}{ \includegraphics*{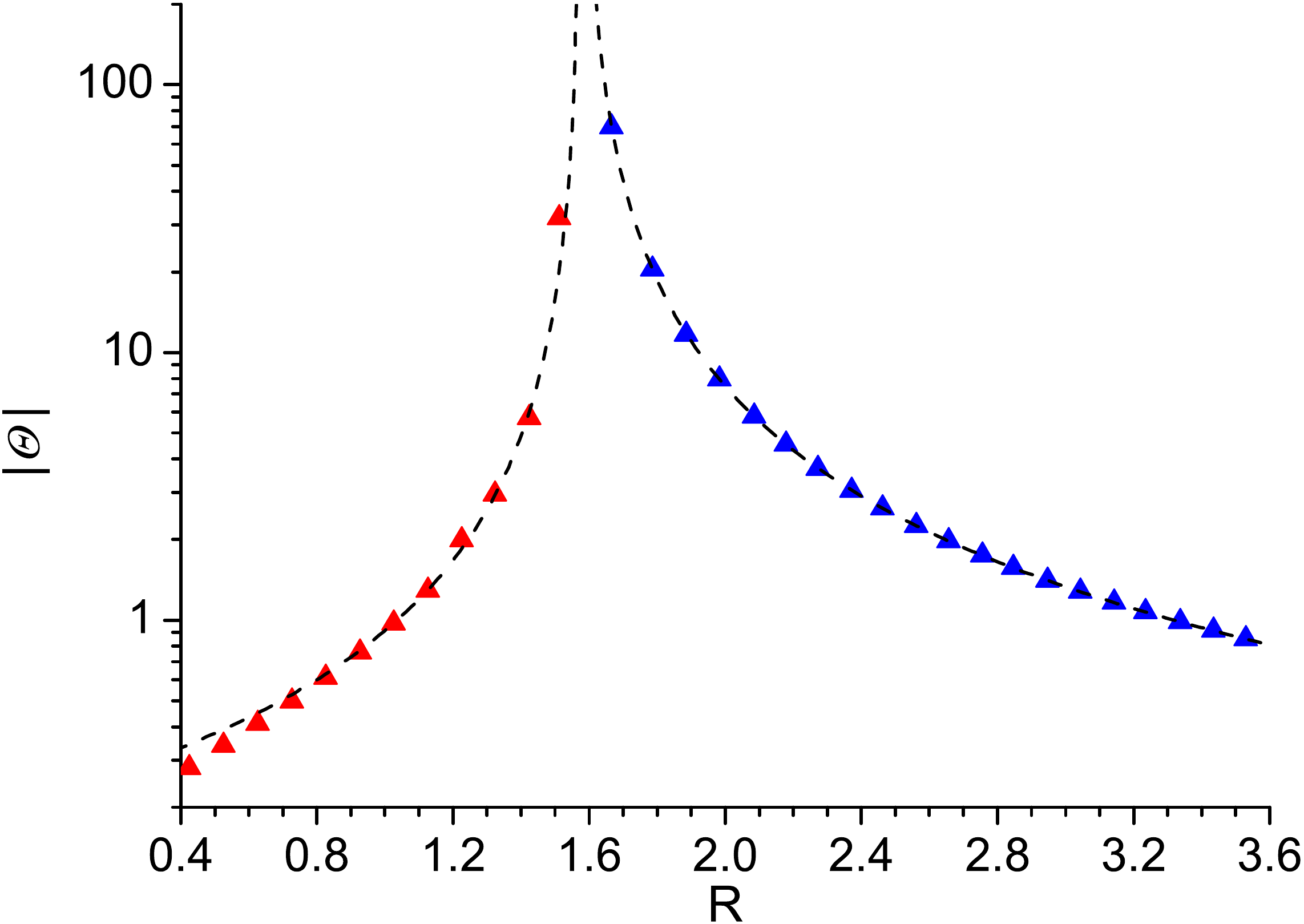}}%
\caption{(Color online) Critical behavior of the statistical temperature. The data of
Fig. \ref{t(re)} are averaged for the Reynolds number intervals $\Delta {\mathrm
{Re}}_{L}=0.1$. The red triangles correspond to the states below ${\mathrm
{Re}}_{L}^{\star}$, and the blue triangles to the states above ${\mathrm
{Re}}_{L}^{\star}$ (${\mathrm {Re}}_{L}^{\star}=1.578$). The dashed curves show the best
fits to the power functions $\Theta=0.42({\mathrm {Re}}^{\star}_{L} - {\mathrm
{R}})^{-1.42}$ (the red triangles) and $\Theta=-2.2({\mathrm {R}} - {\mathrm
{Re}}_{L}^{\star})^{-1.42}$ (the blue triangles).} \label{t_crit(re)}
\end{figure}

Following Onsager \cite{Onsager49}, let us consider local temperatures of flow states,
which in our case will be related to the coherent structures of different size
\begin{equation}\label{eq5}
1/T_{i}=\Delta S^{\mathrm{cum}}_{i}/\Delta E^{\mathrm{cum}}_{i}
\end{equation}
where $E^{\mathrm{cum}}_{i}=\sum_{1}^{i}E_{k}$ and
$S^{\mathrm{cum}}_{i}=\sum_{1}^{i}S_{k}$ are the "cumulative" distributions of the energy
and entropy, respectively, and the increment is calculated as $\Delta
X_{i}=X_{i}-X_{i-1}$. The quantities $E^{\mathrm{cum}}_{i}$ and $S^{\mathrm{cum}}_{i}$
play a role of the energy $E$ and entropy $S$ in the Onsager theory \cite{Onsager49}. In
particular, the energies are similar in two important respects. First,
$E^{\mathrm{cum}}_{i}$ increases as the coherent structures of larger size are included
into consideration, similar as $E$ increased because of clustering of vortices of the
same size in the Onsager theory. Secondly, since the Hamiltonian for the system of
two-dimensional point vortices in inviscid fluid is the part of the kinetic energy of the
system that depends on the relative positions of the vortices (Batchelor
\cite{Batchelor}), the total energy $E$ considered by Onsager represents the kinetic
energy, i.e., similar to what we assumed for the total energy of the system of coherent
structures (Sect. II). The difference with the Onsager theory is that in our case,
because we consider a viscous fluid, the temperature distribution also depends on the
Reynolds number ${\mathrm{Re}}_{L}$ as on a parameter. Taking into account Eqs.
(\ref{eq1}), (\ref{eq3}) and (\ref{eq3_2}), we have
\begin{equation}\label{eq6}
E^{\mathrm{cum}}_{i}=A\sum_{k=1}^{k=i}\tilde{M}_{k}q_{k}^{\gamma}
\end{equation}
and
\begin{equation}\label{eq7}
S^{\mathrm{cum}}_{i}=\ln \Gamma^{\mathrm{cum}}_{i}=-\sum_{k=1}^{k=i}\tilde{M}_{k}(q_{k}
\ln q_{k}+\ln \tilde{M}_{k}-1)+B
\end{equation}
where $A=n^\gamma$ and $B=N\ln(N/n)+N/n$. Then, assuming $N$ and $n$ to be fixed, so that
the Reynolds number determined by Eq. (\ref{eq4}) is fixed too, Eq. (\ref{eq5}) gives
\begin{equation} \label{eq8}
\frac{1}{T_{i}}=-\frac{\tilde{M}_{i}(q_{i} \ln q_{i}+\ln \tilde{M}_{i}-1)}{A
\tilde{M}_{i}q_{i}^{\gamma}}=-\frac{\alpha q_{i}+\beta q_{i}^{\gamma}-1}{A
q_{i}^{\gamma}}
\end{equation}
To represent the temperature in the subsequent Figs. \ref{t(q)}-\ref{t_crit(re)}, the
constant $A$ is set to unity.

Examination of the temperature distributions given by Eq. (\ref{eq8}) shows that they are
characteristically different in different ranges of variation of the Reynolds number.
Specifically, the distributions of three types are observed, which are depicted in Fig.
\ref{t(q)}. At small ${\mathrm {Re}}_{L}$ the temperature is positive for all $q_i$ (the
red curve), at large ${\mathrm {Re}}_{L}$ it is negative for all $q_i$ (the blue curve),
and at intermediate values of ${\mathrm {Re}}_{L}$, the temperature abruptly changes from
positive to negative at some $q{_i}=q{_i}^{\mathrm{c}}$, which corresponds to the maximum
of the entropy (the black curve). In the latter case, the behavior of the temperature is
qualitatively similar to what was predicted by Onsager \cite{Onsager49}. The essential
difference is that in the present case the distribution is dependent upon the Reynolds
number as on a parameter; in particular, the value of $q{_i}^{\mathrm{c}}$ changes with
${\mathrm {Re}}_{L}$, although not regularly because different combinations of parameters
$\alpha$ and $\beta$ entering Eq. (\ref{eq8}) can lead to very close values of ${\mathrm
{Re}}_{L}$ (as, e.g., Fig. \ref{q(re)} evidences).

To see how the temperature of state changes with the Reynolds number, let us introduce
average temperatures $\langle T\rangle$ which are characteristic of different ${\mathrm
{Re}}_{L}$. If the temperature is positive or negative for all $q_i$, as for ${\mathrm
{Re}}_{L} \ll {\mathrm {Re}}_{L}^{\star}$ and ${\mathrm {Re}}_{L} \gg {\mathrm
{Re}}_{L}^{\star}$ in Fig. \ref{t(q)}, it was calculated as an overall average $\langle
T\rangle=\sum_{i} T_{i}\tilde{M}_{i}/\sum_{i} \tilde{M}_{i}$, where $\tilde{M}_{i}$ is
the most probably size distribution of the structures, Eq. (\ref{eq3}). Otherwise, if the
sign of the temperature changes, as for ${\mathrm {Re}}_{L} \sim {\mathrm
{Re}}_{L}^{\star}$ in Fig. \ref{t(q)}, it was calculated for the $T_{i}>0$ and $T_{i}<0$
segments of the dependence $T_{i}=T_{i}(q_{i})$ separately. The results of the
calculations are presented in Fig. \ref{t(re)}. It is seen that the overall average
temperature drastically diverges, changing from positive to negative as the Reynolds
number increases. This divergence of the overall average temperature is accompanied by
large fluctuations of the segment average temperatures in a relatively wide range of the
Reynolds numbers. The state of the flow that has the temperature of the same sign for all
coherent structures, i.e., the overall average temperature, should be associated with a
certain state of the flow, which is laminar or turbulent. Therefore, the laminar and
turbulent states, which correspond to small and large Reynolds numbers, are characterized
by positive and negative temperatures, respectively. Correspondingly, the range of the
Reynolds numbers where the positive and negative temperatures are in mixture should be
associated with the laminar-turbulent transition range.

Figure \ref{t(re)} suggests that the laminar-turbulent transition is similar to a
second-order phase transition \cite{LandauLifshitz_stat_phys} in that at some critical
Reynolds number the temperature drastically diverges, which is accompanied by large
fluctuations of the temperature. A similarity between these two phenomena has been
noticed and discussed in Refs. \onlinecite {Chabaud_Hebral94,TabelingWillaime02,
CortetChiffaudelDaviaudDubrulle10}, where laminar-turbulent transition was studied
experimentally. Chabaud et al. \cite{Chabaud_Hebral94} investigated a jet of gaseous
helium and found that in the transition range the behavior of the width of the velocity
probability density function depending on the size of the spatial structures resembled
the isotherms of a liquid-gas system near the critical point. Tabeling and Willaime
\cite{TabelingWillaime02} studied the flatness of the velocity derivatives in the flow of
gaseous helium between counter-rotating disks (a von K\'{a}rm\'{a}n swirling flow) and
showed that in the transition range the flatness experiences significant changes and has
a power-law dependence on the Reynolds number as $\sim \mathrm{Re}^{0.54}$ below the
critical number $\mathrm{Re_{c}}$ and $\sim (\mathrm{Re}-\mathrm{Re_{c}})^{0.5}$ above
$\mathrm{Re_{c}}$. More recently, Cortet et al. \cite{CortetChiffaudelDaviaudDubrulle10}
investigated the susceptibility of the von K\'{a}rm\'{a}n swirling liquid water flows to
symmetry breaking initiated by different rotation frequencies of the disks and found that
at a critical Reynolds number $\mathrm{Re_{c}}$ the susceptibility diverges as $\chi_{1}
\sim |1/\log{\mathrm{Re}}-1/\log{\mathrm{{Re_{c}}}}|^{-1}$. The quantity $1/{\mathrm {log
{Re}}}$ was interpreted as a temperature due to Castaing \cite{Castaing96}, who
introduced the turbulent flow temperature as a quantity that is conserved along the
length scales in the energy cascade.

The present model shows a similar qualitative behavior of the state temperature as a
function of the Reynolds number ${\mathrm {Re}}_{L}$. To determine critical exponents, we
averaged the data of Fig. \ref{t(re)} by dividing the Reynolds numbers into bins of
length 0.1 (all temperatures were taken into account, not only those shown in Fig.
\ref{t(re)}). In each bin, the mean values of the Reynolds number ($\mathrm {R}$) and the
temperature ($\Theta$) were calculated. The obtained dependence of $\Theta$ on $\mathrm
{R}$ was fitted to the function $\Theta_{\mathrm {theor}}=a|{\mathrm {R}}-{\mathrm
{Re}}_{L}^{\star}|^b$ separately for ${\mathrm {R}} < {\mathrm {Re}}_{L}^{\star}$ and
${\mathrm {R}} > {\mathrm {Re}}_{L}^{\star}$, where $a$ and $b$ are the constants to be
fitted, and ${\mathrm {Re}}_{L}^{\star}$ is an expected critical Reynolds number (Fig.
\ref{t_crit(re)}). The constant $b$ plays a role of the critical exponent. To determine
constants $a$ and $b$ for a selected value of ${\mathrm {Re}}_{L}^{\star}$, the
functional $Q=\sum [(\Theta_{i}-\Theta_{{\mathrm {theor}},i})/\sigma_i]^2$ was evaluated,
where $i$ is the number of the bin, and $\sigma_i$ is the standard deviation of
$\Theta_{i}$ in $i$ bin. The functional was minimized with respect to $a$ and $b$ in a
close vicinity of ${\mathrm {Re}}_{L}^{\star}$ ($|{\mathrm {R}}-{\mathrm
{Re}}_{L}^{\star}|\leq 0.5$). To find an optimal value of ${\mathrm {Re}}_{L}^{\star}$,
the procedure was repeated for different values of ${\mathrm {Re}}_{L}^{\star}$ that
increased from 1.5 to 1.9. The value of ${\mathrm {Re}}_{L}^{\star}$ was considered to be
optimal if the both final values of $Q$ for ${\mathrm {R}}<{\mathrm {Re}}_{L}^{\star}$
and ${\mathrm {R}} > {\mathrm {Re}}_{L}^{\star}$ were smaller than the corresponding
values of $Q$ for the other values of ${\mathrm {Re}}_{L}^{\star}$. The best fit was
achieved at ${\mathrm {Re}}_{L}^{\star} =1.578$, in which case the data are approximated
by the functions $\Theta=0.42({\mathrm {Re}}^{\star}_{L} - {\mathrm {R}})^{-1.42}$ at
${\mathrm {R}} < {\mathrm {Re}}_{L}^{\star}$ and $\Theta=-2.2({\mathrm {R}} - {\mathrm
{Re}}_{L}^{\star})^{-1.42}$ at ${\mathrm {R}} > {\mathrm {Re}}_{L}^{\star}$ (Fig.
\ref{t_crit(re)}). The fluctuations of the temperature sharply increase toward the
critical Reynolds number (exponentially rather than by a power law) and reach
$\sigma/\Theta \sim 1$ and $\sigma/\Theta \sim 10$ at ${\mathrm {R}} < {\mathrm
{Re}}_{L}^{\star}$ and ${\mathrm {R}} > {\mathrm {Re}}_{L}^{\star}$, respectively.

The observed behavior of the temperature with the Reynolds number is similar to
second-order phase transitions in several essential respects: the temperature exhibits a
power-law divergence at the critical point (the critical Reynolds number), the critical
exponents below and above the critical point are practically equal to each other and have
the same order of magnitude as those in second-order phase transitions, and the
fluctuations of the temperature in the vicinity of the critical point are as large or
larger than the mean value of the temperature (Stanley \cite{Stanley}). At the same time,
drawing a direct analogy to second-order phase transitions is problematic, primarily
because the flow cannot be divided into parts without a loss of its essential properties
(Sect. II). Consequently, the system of particles representing the flow does not have the
property of self-similarity, which plays a key role in second-order phase transitions and
critical phenomena (Kadanoff \cite{Kadanoff66} and Wilson \cite{Wilson75}).

The drastic change of the state temperature at ${\mathrm {Re}}_{L} \sim {\mathrm
{Re}}_{L}^{\star}$ offers a criterion to distinguish between laminar and turbulent states
that is more definite than the dependence of $\langle q\rangle$ upon ${\mathrm {Re}}_{L}$
(Fig. \ref{q(re)}). The unreasonably low value of ${\mathrm {Re}}_{L}^{\star}$ ($\approx
1.6$) we obtained is, in a considerable degree, a result of the flexibility in the
definition of the size of the elementary volume. Equation (\ref{eq4}) assumed that it was
the product of the Kolmogorov length and velocity dissipation microscales $\eta
v_{\eta}=\nu$. However, the space scale must not be necessarily equal to $\eta$. For
example, the scale dividing the dissipation and inertial subranges $l_{\mathrm{DI}}=60
\eta$  (Pope \cite{Pope00}) can be taken as the space scale. Correspondingly, because in
the inertial subrange $v \sim (\epsilon l)^{1/3}$, $v_{\mathrm{DI}}=(\epsilon
l_{\mathrm{DI}})^{1/3}$ can serve as the velocity scale. Then, the value of the critical
Reynolds number ${\mathrm {Re}}_{L}^{\star}$ increases $60^{4/3}$ times, which leads to a
more reasonable value of ${\mathrm {Re}}_{L}^{\star} \approx 4 \times 10^2$.

It would be of interest to determine the statistical temperature of the flow by
experiment or numerical simulations and test the predicted behavior of the temperature
with the Reynolds number. For this, Eq. (\ref{eq8}) can be rewritten as $1/T_{i} \sim
-(q_{i} \ln q_{i}+\ln{M}_{i}-1)/E_{i}$, where $E_{i}$ is the kinetic energy of $i$
structure, $M_{i}$ is the size distribution of the structures, and $q_{i}=V_{i}/\eta^3$,
where $V_{i}$ is the volume of $i$ structure, and $\eta$ is the Kolmogorov microscale.
According to this equation, all one needs to know to calculate the temperature is the
volumes of the coherent structures, their kinetic energies, and the size distribution of
the structures. Such information can be obtained, in principle, by direct numerical
simulations, similar to those of Moisy and Jim\'{e}nez \cite{MoisyJimenez04}, in which
the size distribution of the coherent structures was determined and the kinetic energies
of the structures could be calculated. of the structures could be calculated.

\section{Conclusions} Using the previously developed statistical model of laminar/turbulent
states of a fluid flow, which considered the flow as a collection of coherent structures
of various size \cite{Chekmarev13}, a statistical temperature of the flow states has been
determined that depends on the Reynolds number. It has been shown that at small Reynolds
numbers, associated with laminar states, the temperature is positive, while at large
Reynolds numbers, associated with turbulent states, it is negative. At intermediate
Reynolds numbers, the temperature abruptly changes from positive to negative as the
structure size increases, similar to what Onsager predicted for a system of parallel
point-vortices in an inviscid fluid \cite{Onsager49}. It has also been shown that in the
range of intermediate Reynolds numbers, which is associated with the transition between
laminar and turbulent states, the temperature drastically diverges, exhibiting a
power-law behavior characteristic of second-order phase transitions. The critical
exponents below and above critical Reynolds number are close to each other and have the
same order of magnitude as those in the second-order phase transitions.

\section{Acknowledgments}
I thank R. Khairulin for useful discussions of the critical phenomena.

\newpage

\end{document}